\documentclass[prl, twocolumn, superscriptaddress]{revtex4-1}
\usepackage{bm, amsmath, amsfonts, amssymb, braket}
\usepackage{times}
\usepackage{multirow}
\usepackage[dvipdfmx]{graphicx}
\usepackage{float, color}

\usepackage[
pagebackref=false,
colorlinks=true,
linkcolor=blue,
urlcolor=blue,
filecolor=black,
citecolor=red,
pdfstartview=FitV,
pdftitle={},
pdfauthor={},
pdfsubject={},
pdfkeywords={},
pdfpagemode=None,
bookmarksopen=true
]{hyperref}

\begin{document}

\newcommand{\ii}{\text{i}}
\newcommand{\SE}{S_{E}}
\newcommand{\SBE}{S_{E}^{\rm boundary}}
\newcommand{\HB}{H^{\rm boundary}}
\newcommand{\WEone}{W_{1} (E)}
\newcommand{\WEthree}{W_{3} (E)}
\newcommand{\jEone}{j (x, E)}
\newcommand{\jEthree}{\bm{j} (\bm{x}, E)}
\newcommand{\jAEthree}{\bm{j}^{A} (\bm{x}, E)}
\newcommand{\NRE}{N_{\rm R} (E)}
\newcommand{\NLE}{N_{L} (E)}

\newcommand{\phase}{\phi}
\newcommand{\texture}{\Theta}

\title{Topological Field Theory of Non-Hermitian Systems}

\author{Kohei Kawabata}
\email{kawabata@cat.phys.s.u-tokyo.ac.jp}
\affiliation{Department of Physics, University of Tokyo, 7-3-1 Hongo, Bunkyo-ku, Tokyo 113-0033, Japan}

\author{Ken Shiozaki}
\email{ken.shiozaki@yukawa.kyoto-u.ac.jp}
\affiliation{Yukawa Institute for Theoretical Physics, Kyoto University, Kyoto 606-8502, Japan}

\author{Shinsei Ryu}
\email{shinseir@princeton.edu}
\affiliation{Department of Physics, Princeton University, Princeton, New Jersey, 08540, USA}

\date{\today}

\begin{abstract}
Non-Hermiticity gives rise to unique topological phases without Hermitian analogs. However, the effective field theory has yet to be established.
Here, we develop a field-theoretical description of the intrinsic non-Hermitian topological phases. 
Because of the dissipative and nonequilibrium nature of non-Hermiticity, our theory is formulated solely in terms of spatial degrees of freedom, which contrasts with the conventional theory defined in spacetime. 
It provides the universal understanding about non-Hermitian topological phenomena, such as the unidirectional transport in one dimension and the chiral magnetic skin effect in three dimensions. Furthermore, it systematically predicts new physics; we illustrate this by revealing transport phenomena and skin effects in two dimensions induced by a perpendicular spatial texture.
From the field-theoretical perspective, the non-Hermitian skin effect, which is anomalous localization due to non-Hermiticity, is shown to be a signature of an anomaly.
\end{abstract}

\maketitle
Topology plays a key role in modern physics.
In particular, topological phases of matter have been arguably
one of the most actively-studied condensed-matter systems
in recent years~\cite{HK-review, QZ-review, CTSR-review}.
A universal understanding of topological phases is obtained by
topological field theory in spacetime.
For example, the Chern-Simon theory describes the quantum Hall
effect~\cite{Redlich:1983dv, Niemi:1983rq, Zhang-89, Frohlich:1990xz, Fradkin:1991wy, Wen:1992uk},
and the axion electrodynamics describes the magnetoelectric effect~\cite{QHZ-08, Essin-09}.
One of the consequences of the topological field theory description
is the bulk-boundary correspondence:
in the presence of a boundary,
a topological field theory is gauge dependent at
the boundary, and
this gauge noninvariance must be compensated
by
an anomaly at the boundary~\cite{Callan-85}.

While topological phases were mainly investigated for Hermitian systems at
equilibrium, topological phases of non-Hermitian systems are attracting growing
interest~\cite{Rudner-09, Sato-11, *Esaki-11, Hu-11, Schomerus-13, Longhi-15,
  Lee-16, Leykam-17, Xu-17, Shen-18, *Kozii-17, Takata-18, MartinezAlvarez-18,
  Gong-18, *Kawabata-19, YW-18-SSH, *YSW-18-Chern, Kunst-18, McDonald-18,
  Lee-Thomale-19, Liu-19, Lee-Li-Gong-19, KSUS-19, ZL-19, Chernodub-20, Hirsbrunner-19,
  Zirnstein-19, Borgnia-19, KBS-19, Yokomizo-19, McClarty-19, Song-19,
  Bergholtz-19, JYLee-19, Yoshida-19, Schomerus-20, Herviou-19-ES, Chang-20,
  Wanjura-20, Zhang-20, OKSS-20, Xi-19, Terrier-20, Xue-20,
  Bessho-20, Denner-20, Ma-20, Okugawa-20, KSS-20}.
Non-Hermiticity arises from dissipation~\cite{Konotop-review,
  Christodoulides-review, BBK-review},
and the interplay between non-Hermiticity and topology
leads to new physics in open classical and quantum systems~\cite{Poli-15,
  Zeuner-15, Weimann-17, Xiao-17, St-Jean-17, Bahari-17, Zhou-18, Harari-18,
  *Bandres-18, Zhao-19, Brandenbourger-19-skin-exp, *Ghatak-19-skin-exp,
  Helbig-19-skin-exp, *Hofmann-19-skin-exp, Xiao-19-skin-exp,
  Weidemann-20-skin-exp, Yamauchi-20}.
One of the remarkable consequences of non-Hermiticity is new types of topological phases without Hermitian analogs.
For example,
non-Hermitian topological phases arise generally in odd spatial dimensions~\cite{Gong-18, KSUS-19},
while no topological phases appear in these dimensions for Hermitian systems without symmetry.
These unique topological phases arise from the complex-valued nature of spectra,
which enables two types of energy gaps, i.e., point and line
gaps~\cite{KSUS-19}.
In the presence of a boundary, such intrinsic non-Hermitian topology leads to
the anomalous localization of an extensive number of eigenstates~\cite{Zhang-20, OKSS-20}---the non-Hermitian skin effect~\cite{Lee-16, YW-18-SSH, Kunst-18}.

However, topological field theory has yet to be established for non-Hermitian systems.
The Chern-Simons theory was shown to remain well defined even for non-Hermitian Chern insulators~\cite{Hirsbrunner-19}.
Still, this theory only describes non-Hermitian topological phases that are continuously deformed to Hermitian phases.
Field-theoretical characterization of intrinsic non-Hermitian topology has
remained elusive, although it is crucial for understanding and exploring non-Hermitian topological phenomena including the skin effect. 

In this Letter, we develop topological field theory of non-Hermitian systems.
We show that field theory of intrinsic non-Hermitian topology is formulated
solely by spatial degrees of freedom
as a consequence of the dissipative and nonequilibrium nature of
non-Hermiticity.
This contrasts with the conventional theory defined by both spatial and temporal
degrees of freedom.
Our theory universally describes and systematically predicts unique non-Hermitian topological phenomena.
Furthermore, we show that the non-Hermitian skin effect is a signature of an anomaly.

\paragraph{Non-Hermitian topology.\,---}
Non-Hermitian systems
give rise to
unique topological phases that have no counterparts in Hermitian systems.
Such intrinsic non-Hermitian topology arises even in one dimension.
Suppose that a non-Hermitian Bloch Hamiltonian $H(k)$
has a point gap,
i.e., 
it is invertible in terms of reference energy $E \in \mathbb{C}$ [i.e., $\det
\left[ H(k) - E \right] \neq 0$ for all $k$]~\cite{Gong-18, KSUS-19}.
Then, the following winding number $\WEone \in \mathbb{Z}$ is well defined:
\begin{equation}
\WEone := - \oint_{\rm BZ} \frac{dk}{2\pi\ii} \left( \frac{d}{dk} \log \det \left[ H( k ) - E \right] \right).
	\label{eq: 1D-winding}
\end{equation}
Notably, $\WEone$ vanishes when $H(k)-E$ is Hermitian. 
This is consistent with the absence of topological phases in one-dimensional Hermitian systems without symmetry~\cite{HK-review, QZ-review, CTSR-review}.

A prototypical lattice model with nontrivial $W_{1}$ is given by~\cite{Hatano-Nelson-96, *Hatano-Nelson-97}
\begin{equation}
\hat{H} = -\frac{1}{2} \sum_{n} \left[ \left( 1+\gamma \right) \hat{c}_{n+1}^{\dag} \hat{c}_{n} + \left( 1-\gamma\right) \hat{c}_{n}^{\dag} \hat{c}_{n+1} \right],
	\label{eq: Hatano-Nelson}
\end{equation}
where $\hat{c}_{n}$ ($\hat{c}_{n}^{\dag}$) annihilates (creates) a particle on
site $n$, and $\gamma \in \mathbb{R}$ denotes the asymmetry of the hopping
amplitudes and describes the degree of non-Hermiticity.
The corresponding Bloch Hamiltonian reads $H( k ) = - \cos k + \ii \gamma \sin k$
and winds around the origin in the complex energy plane.
Consequently, we have $\WEone = \mathrm{sgn} \left( \gamma \right)$ as long as the reference energy $E$
is inside the region surrounded by the loop of $H( k )$.
Despite the presence of a point gap, an energy gap for the real part of the
spectrum closes at $k = \pm \pi/2$, i.e., $\mathrm{Re}\,H \left( k =\pm \pi/2 \right) = 0$;
this type of energy gap is called a line gap~\cite{Shen-18, KSUS-19}.
To understand a universal feature of non-Hermitian topology, let us consider the
continuum Dirac Hamiltonian around
the
line-gap-closing points:
\begin{equation}
H( k ) = k + \ii \gamma.
	\label{eq: 1D-Dirac}
\end{equation}
This non-Hermitian Dirac Hamiltonian, similarly to its lattice counterpart, is characterized by
the nonzero winding number $\WEone = \mathrm{sgn} \left( \gamma - \mathrm{Im}\,E
\right)/2$.

An important consequence of nontrivial $W_{1}$ is the non-Hermitian skin effect~\cite{Zhang-20, OKSS-20}. In the presence of a boundary, there appear $\left| \WEone \right|$ eigenstates with the eigenenergy $E$ at the boundary. Notably, $\WEone$ can be nontrivial for many choices of the reference energy $E$, and consequently, an extensive number of eigenstates are localized at the boundary. In the lattice model~(\ref{eq: Hatano-Nelson}), all the eigenstates are localized at the right (left) edge for $\gamma > 0$ ($\gamma < 0$). Such anomalous localization is impossible in Hermitian systems and hence a unique non-Hermitian topological phenomenon.

\paragraph{Topological field theory.\,---}
Before developing effective field theory for intrinsic non-Hermitian
topology, let us briefly recall the Hermitian case. 
The effective field theory for Hermitian systems is obtained by introducing a
gauge potential $( \bm{A}, \phi )$ to a microscopic Hamiltonian and
integrating out matter degrees of freedom.
The quantum partition function of a
Hamiltonian $H(\bm{k})$
is given by path integral
as $Z[ \bm{A}, \phi] = \int \mathcal{D}\bar{\psi}\mathcal{D}\psi\,e^{\ii {\cal S}}$
with the (real-time) action
\begin{equation}
{\cal S} = \int \bar{\psi} \left[ \ii \partial_{t} + \phi - H( -\ii \partial_{\bm{x}} - \bm{A}) \right] \psi\,d^{d}x dt.
	\label{eq: action-spacetime}
\end{equation}
Here, $\psi$ and $\bar{\psi}$ describe matter degrees of freedom.
Integrating over the matter field,
we obtain
the effective action $S[ \bm{A}, \phi]$
for the external field,
$e^{\ii S[\bm{A},\phi]} := Z[ \bm{A}, \phi ]/Z [0]$
with 
\begin{equation}
  Z[ \bm{A}, \phi ] = \det \left[ \ii \omega + \phi
    - H( \bm{k} - \bm{A}) \right].
	\label{eq: partition-function-spacetime}
\end{equation} 
In the presence of an energy gap,
the effective action is
given by a local functional of
$(\bm{A},\phi)$.
The response of the system to the external field
can be read off from the current density
$\bm{j} := \delta S/\delta \bm{A}$.
In this formulation, the topological invariant
that appears in the topological term of the effective action
is given by 
the Green's function
\cite{Ishikawa:1987zi}
\begin{equation}
  G_{0}( \bm{k}, \omega ) := \left( \ii \omega - H( \bm{k}) \right)^{-1}, 
	\label{eq: Green-spacetime}
\end{equation}
which is a non-Hermitian matrix.
For example,
if we consider the Dirac Hamiltonian $H( \bm{k}) = k_{x} \sigma_{x} + k_{y} \sigma_{y} + m \sigma_{z}$ with Pauli matrices $\sigma_{i}$'s, we obtain the $\left( 2+1 \right)$-dimensional Chern-Simons theory, which describes the quantum Hall effect.

The above path integral, in its Euclidean version,
assumes the Gibbs state as an equilibrium density matrix.
On the other hand, 
for the non-Hermitian case, 
the thermal equilibrium is no longer achieved, and
it is generally unclear
what kind of path integral one should set up.
This constitutes a fundamental difficulty for developing effective field
theory.
This may be tackled, for example, by the Schwinger-Keldysh approach~\cite{Kamenev-textbook}.
Nevertheless,
as long as an energy gap for the real part of the spectrum (i.e., line gap)
stays open, the above procedure gives rise to
topological field theory even for non-Hermitian systems.
In this case, non-Hermitian topological phases are continuously
deformable to their Hermitian counterparts~\cite{KSUS-19},
and share the same topological field theory.
For example, for non-Hermitian Chern insulators,
the above procedure delivers the $( 2+1 )$-dimensional Chern-Simons theory
\cite{Hirsbrunner-19}.
However, this is not the case for intrinsic non-Hermitian topology.
For the non-Hermitian Dirac Hamiltonian (\ref{eq: 1D-Dirac}), the line gap vanishes,
and matter degrees of freedom cannot be integrated out
safely; if we calculate
$S[ \bm{A}, \phi]$ from Eq.~(\ref{eq: partition-function-spacetime}),
it 
is ill defined.
We also note that
the quantization of $\WEone$ in Eq.~(\ref{eq: 1D-winding})
is guaranteed by the point gap $E( k) \neq E$
instead of the line gap $\mathrm{Re}\,E ( k) \neq E$,
which is a unique gap structure due to the complex-valued nature of the spectrum~\cite{Gong-18, KSUS-19, KBS-19}.

We thus seek a different formulation of field theory
for intrinsic non-Hermitian topological phases.
Since these phases arise out of equilibrium,
the temporal degree of freedom should play a special role. Then, let us Fourier
transform the field operator in time by
$\psi( {\bm x}, t) = \int \psi_{E}({\bm x}) e^{-\ii Et} dE$ 
and focus on fixed energy $E$.
We also switch off the scalar potential $\phi$ and focus on
a time-independent vector potential
${\bm A}({\bm x})$.
The action (\ref{eq: action-spacetime}) in spacetime reduces to
\begin{equation}
{\cal S}_E = \int \bar{\psi}_{E} \left[ H ( -\ii \partial_{\bm{x}} - \bm{A} ) - E \right] \psi_{E}\,d^{d}x.
\end{equation}
In contrast to Eq.~(\ref{eq: action-spacetime}),
this action is written solely
in terms of the spatial degrees of freedom.
The functional integral
\begin{equation}
  Z_E[ \bm{A}]
  =
  \int \mathcal{D}\bar{\psi}_E \mathcal{D}\psi_E\, e^{ \ii {\cal S}_E}
  = \det \left[ H( \bm{k} - \bm{A}) - E \right]
	\label{eq: partition-function-space}
\end{equation} 
serves as a generating functional of
the single-particle Green's function $\left( E - H ( \bm{k} ) \right)^{-1}$
with reference energy $E$.
It is therefore expected to capture all physical information---including topological one such as the non-Hermitian skin effect.
This type of spatial field theory
is commonly used 
for Anderson localization~\cite{Efetov-textbook, Altland-Simons-textbook}
and also for Hermitian topological systems in odd dimensions~\cite{Schnyder-08}.
It is discussed also
for Floquet systems and their boundary unitary operators~\cite{Glorioso-19}.

To further emphasize the special role played by the temporal direction, 
we note that 
one of the wavenumbers, such as 
$k$ in Eq.~(\ref{eq: 1D-Dirac}), plays a similar role to frequency
$\omega$ for Hermitian systems; the inverse of the Green's function $G_{0}^{-1} ( \bm{k}, \omega)$ in Eq.~(\ref{eq: Green-spacetime}) for a Hermitian Hamiltonian is identified with a non-Hermitian Hamiltonian $H ( \bm{k} )$ in Eq.~(\ref{eq: partition-function-space}) by replacing $\omega$ with $k$. Thus, the effective action of non-Hermitian systems in $d+0$ dimensions is mathematically equivalent to that of Hermitian systems in $\left( d-1 \right) + 1$ dimensions.
Consistently, $d$-dimensional non-Hermitian systems are topologically classified in the same manner as $\left( d-1 \right)$-dimensional Hermitian systems in the same symmetry class~\footnote{Non-Hermitian Bloch Hamiltonians $H \left( \bm{k} \right)$ in $d$ dimensions with Altland-Zirnbauer symmetry have the same topological classification as Hermitian Bloch Hamiltonians in $d-1$ dimensions with the same symmetry class~\cite{KSUS-19}, in which time-reversal and particle-hole symmetry is respectively defined by $\mathcal{T} H^{*} \left( \bm{k} \right) \mathcal{T}^{-1} = H \left( -\bm{k} \right)$ and $\mathcal{C} H^{T} \left( \bm{k} \right) \mathcal{C}^{-1} = - H \left( -\bm{k} \right)$ with unitary matrices $\mathcal{T}$ and $\mathcal{C}$ (see also Appendix~E in Ref.~\cite{KSUS-19}).}; the difference of one dimension corresponds to time~\footnote{Reference~\cite{Xi-19} argued that topological classification of non-Hermitian systems in one dimension is the same as the Hermitian case from the field-theoretical perspective. At the face value, this result may contradict the intrinsic non-Hermitian topological phases in one dimension~\cite{Gong-18, KSUS-19}. However, since Ref.~\cite{Xi-19} assumes a line gap and a spacetime formulation, it is compatible with our space formulation for a point gap.}.
The degree of a point gap, such as $\gamma$ in Eq.~(\ref{eq: 1D-Dirac}), gives a relevant energy scale and ensures the local expansion of the effective action by the gauge potential.

\paragraph{One dimension.\,---}

Below, we explicitly provide field theories of intrinsic non-Hermitian topology and discuss unique phenomena including the skin effect.
For the non-Hermitian Dirac Hamiltonian~(\ref{eq: 1D-Dirac}), the effective action is
\begin{equation}
\SE [ A ]
\simeq \ii\,\mathrm{tr} \left[ ( H ( -\ii \partial_{x} ) - E )
  A( x ) \right],
\end{equation}
where the vector potential $A$ is assumed to be sufficiently small.
After taking the sum explicitly,
this further reduces to
\begin{equation}
\SE [ A ] 
= \WEone \int A( x ) dx,
	\label{eq: TQFT-1D}
\end{equation}
where the winding number $\WEone$ is defined for reference energy $E$ as Eq.~(\ref{eq: 1D-winding}).
This is the $\left( 1+0\right)$-dimensional Chern-Simons theory. As discussed above, replacing $x$ with $t$, we have the $\left( 0+1 \right)$-dimensional Chern-Simons theory, which describes Hermitian systems in zero dimension.

From this effective action, the current is obtained as
\begin{equation}
\jEone := \frac{\delta \SE [ A ]}{\delta A( x)} = \WEone.
	\label{eq: current-1D}
\end{equation}
Thus, particles unidirectionally flow from the left to the right (from the right
to the left) for $W_{1} > 0$ ($W_{1} < 0$).
Consistently, in the lattice model~(\ref{eq: Hatano-Nelson}),
the hopping amplitude from the left to the right is greater (smaller) than that
from the right to the left for $W_{1} > 0$ ($W_{1} < 0$).
This type of directional amplification ubiquitously appears, for example, in
open photonic systems~\cite{Longhi-15, Gong-18, Wanjura-20, Xue-20}
and parametrically-driven bosonic systems~\cite{McDonald-18}.
The topological field theory (\ref{eq: TQFT-1D}) underlies such unidirectional transport induced by asymmetric hopping.

\paragraph{Skin effect as an anomaly.\,---}
In the presence of a boundary,
the topological action is no longer gauge invariant.
Suppose that the system described by Eq.~(\ref{eq: TQFT-1D}) lies in $x_{\rm L}
\leq x \leq x_{\rm R}$, outside of which is the vacuum.
Then,
under the gauge transformation $A \to A + df/dx$ with an arbitrary function $f$,
the effective action $\SE$ changes into
$\SE + \WEone \left[ f( x_{\rm R} ) - f ( x_{\rm L} ) \right]$ and explicitly depends on the choice of the gauge $f$. To retain gauge invariance, additional degree of freedom is required at the boundary $x = x_{\rm L}, x_{\rm R}$. This boundary system reads
\begin{equation}
\SBE = - \WEone \left[ \varphi( x_{\rm R}) - \varphi ( x_{\rm L} ) \right],
	\label{eq: TQFT-1D-boundary}
\end{equation}
where $\varphi( x )$ denotes the phase of the wavefunction at $x$. Since $\varphi$ changes to $\varphi + f$ through the gauge transformation, $\SBE$ changes into $\SBE - \WEone \left[ f ( x_{\rm R} ) - f ( x_{\rm L} ) \right]$. Thus, while $\SE$ and $\SBE$ are individually gauge dependent, their combination $\SE + \SBE$ is indeed gauge invariant.

The boundary action (\ref{eq: TQFT-1D-boundary})
describes a pair of the charges $\WEone$ and $-\WEone$ at $x = x_{\rm R}$ and $x =
x_{\rm L}$, respectively. These charges correspond to skin modes.
Importantly, reference energy $E$ can be chosen arbitrarily as long as $\WEone$
is nontrivial. An extensive number of the charges appear at the boundary, which
correspond to an extensive number of skin modes.
Thus, the skin effect originates from a non-Hermitian anomaly.
This contrasts with Hermitian systems in one dimension,
in which an anomaly results in only a finite number of symmetry-protected zero-energy modes at the boundary.
We note that an anomaly discussed here is distinct from a dynamical anomaly in Refs.~\cite{JYLee-19, Terrier-20, Bessho-20, Okuma-21}.

\paragraph{Three dimensions.\,---}
Topological field theories
are formulated also in higher dimensions.
In general, non-Hermitian systems in $d$ dimensions are described by the $\left( d+0 \right)$-dimensional Chern-Simons theory 
for odd $d$.
This contrasts with Hermitian systems, which are described by the $\left( d+1 \right)$-dimensional Chern-Simons theory 
for even $d$.

In three dimensions, for example, the non-Hermitian Dirac Hamiltonian
$H ( \bm{k} ) = k_{x} \sigma_{x} + k_{y} \sigma_{y} + k_{z} \sigma_{z} + \ii \gamma$ results in the $\left( 3+0 \right)$-dimensional Chern-Simons theory:
\begin{equation}
  \SE [ \bm{A} ] = \frac{\WEthree}{4\pi} \sum_{ijk} \int \varepsilon_{ijk}
  A_{i}( \bm{x} ) \partial_{j} A_{k}( \bm{x} )d^{3}x,
	\label{eq: TQFT-3D}
\end{equation}
where $W_{3}$ is the three-dimensional winding number. The current density of this theory is
\begin{equation}
\jEthree := \frac{\delta \SE [ \bm{A}]}{\delta \bm{A}(\bm{x})}
= \frac{\WEthree}{2\pi} \bm{B}( \bm{x}),
	\label{eq: current-3D}
\end{equation}
where $\bm{B} := \nabla \times \bm{A}$ is a magnetic field.
Thus, particles flow along the direction of the magnetic field $\bm{B}$. 
This is the chiral magnetic effect~\cite{Fukushima-08} in which non-Hermiticity induces chirality imbalance in a similar manner to an electric field.
This further means that the direction of amplification can be controlled by a
magnetic field, which is a unique property of three-dimensional systems. It is
also remarkable that Ref.~\cite{Bessho-20} recently constructed a lattice model
that exhibits the non-Hermitian chiral magnetic effect.
This Letter gives its field-theoretical understanding.

Under the open boundary conditions, $\SE$ is gauge dependent.
For the quantum Hall effect,
which is described by the $\left( 2+1 \right)$-dimensional Chern-Simons theory,
the gauge noninvariance is balanced
with an anomaly of chiral fermions at the boundary~\cite{Callan-85}.
In the non-Hermitian case, 
the boundary degrees of freedom are described by
a Hamiltonian with a single exceptional point, 
$H( \bm{k}) = \pm k_{x} - \ii k_{y}$, for $\left| \WEthree \right| = 1$~\cite{Denner-20, supplement}, which reduces to the inverse of the Green's function of the conventional chiral fermions by replacing $k_{y}$ with frequency $\omega$. In $1+1$ dimensions, a chiral anomaly is described by $\partial_{x} j_{x}^{A} + \partial_{t} j_{t}^{A} = E/\pi$ with an axial current $\left( j_{x}^{A}, j_{t}^{A} \right)$ and an electric field $E := \partial_{x} A_{t} -\partial_{t} A_{x}$~\cite{Adler-69, Bell-69, Peskin-textbook}.
Replacing time with another spatial component $y$, we have the following non-Hermitian analog of the anomaly equation~\cite{supplement}:
\begin{equation}
\nabla \cdot \jAEthree = \frac{\WEthree B ( \bm{x} )}{\pi},
\end{equation}
where $B := \partial_{x} A_{y} - \partial_{y} A_{x}$ is a magnetic field
perpendicular to the surfaces. In terms of the global quantities such as the
charge
$N_{\rm R}$ ($N_{\rm L}$) of the right-moving (left-moving)
particle $H( \bm{k} ) = k_{x} - \ii k_{y}$
[$H ( \bm{k} ) = -k_{x} - \ii k_{y}$], as well as the magnetic flux
$\Phi := \int B ( \bm{x} ) d^{2}x$, this anomaly equation reduces to
\begin{equation}
\NRE - \NLE = \frac{\WEthree \Phi}{\pi}.
    \label{eq: 3D-anomaly-skin}
\end{equation}
Combining it with the global conservation law $N_{\rm R} + N_{\rm L} = 0$ due to $\mathrm{U}( 1 )$ symmetry, we have $N_{\rm R} = W_{3}\Phi/2\pi$ and $N_{\rm L} = - W_{3}\Phi/2\pi$. Here, $\Phi/2\pi$ is the number of the fluxes since $2\pi$ denotes the flux quantum in the natural units (i.e., $e = \hbar = 1$). 
Thus, a signature of the topological action
in three dimensions appears as the skin effect induced by a magnetic field. The
number of the skin modes is given by the topological invariant $W_{3}$ and the
number of fluxes. 
While Ref.~\cite{Bessho-20} predicted this three-dimensional version of the skin effect---chiral magnetic skin effect---on the basis of the bulk topological invariant, we here derive it from a chiral anomaly at a boundary.
It occurs also in a lattice model~\cite{supplement, Nakamura-20}. 

\paragraph{Two dimensions.\,---}
For Hermitian systems, topological field theories of superconductors in $0+1$ and $1+1$ dimensions, and insulators in $2+1$ and $3+1$ dimensions are derived from the Chern-Simons theories in $2+1$ and $4+1$ dimensions, respectively~\cite{QHZ-08}.
Topological field theories of non-Hermitian systems in even dimensions are also derived from higher-dimensional ones. For example, let us reduce the $z$ direction of the $\left( 3+0 \right)$-dimensional theory~(\ref{eq: TQFT-3D}) by considering $z$ to be a parameter and making the gauge potential $\bm{A}$ be independent of $z$.
Then, the effective action~(\ref{eq: TQFT-3D}) reduces to
\begin{equation}
\SE [ \bm{A} ]
  = \frac{1}{2\pi} \sum_{ij} \int \theta ( \bm{x}, E ) \varepsilon_{ij}
      \partial_{i} A_{j} ( \bm{x}) d^{2}x
	\label{eq: TQFT-2D}
\end{equation}
with the Wess-Zumino term $\theta$~\cite{WZ-71, supplement}.
This is the effective action of non-Hermitian systems in two dimensions.
Here, $\theta$ is a non-Hermitian analog of the electric polarization in $\left( 1+1 \right)$-dimensional Hermitian systems~\cite{Vanderbilt-textbook}, and $\mathbb{Z}_{2}$ quantized in the presence of reciprocity or particle-hole symmetry.

\begin{figure}[t]
\centering
\includegraphics[width=86mm]{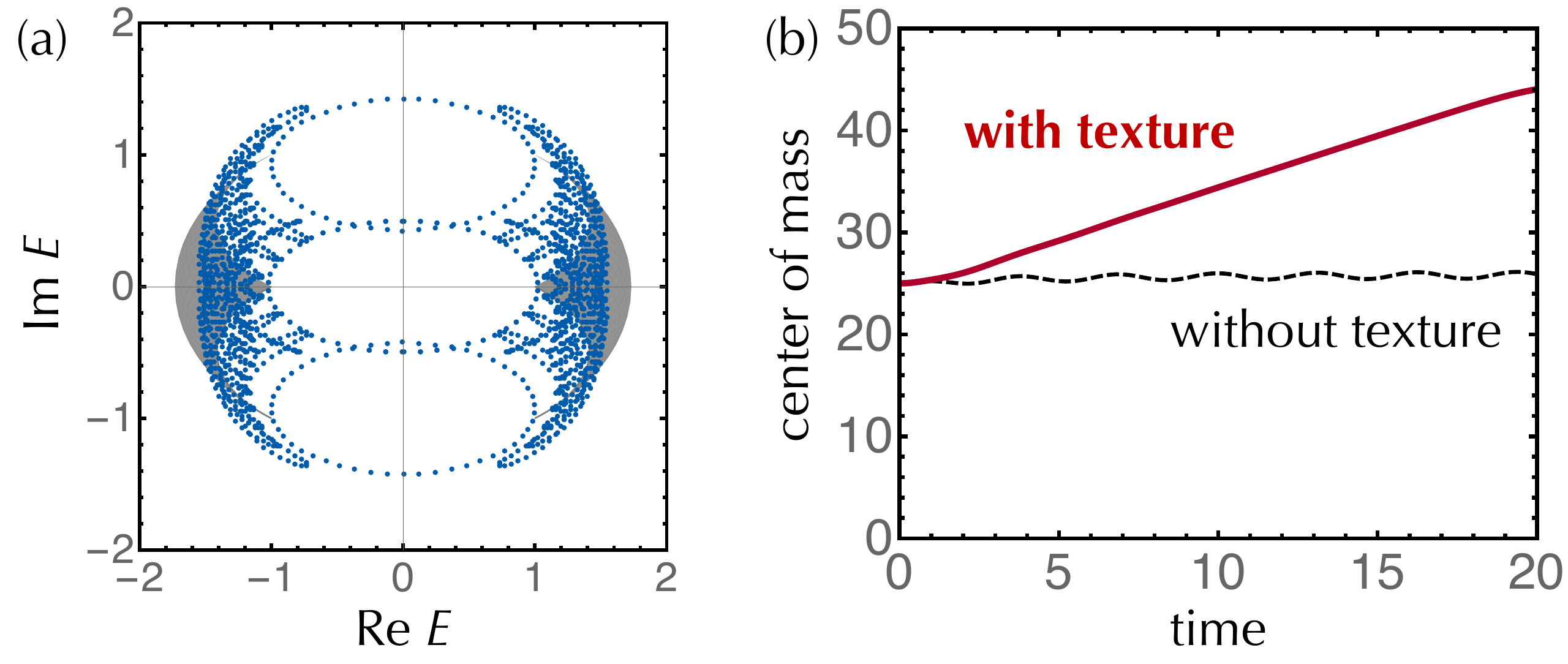} 
\caption{Non-Hermitian topological phenomena in two dimensions. The periodic boundary conditions are imposed along both directions ($L_{x} = 50, L_{y} = 20$; $\gamma = 0.5$). (a)~Complex spectra in the presence ($\texture =1$, blue dots) and absence ($\texture = 0$, gray regions) of the spatial texture. (b)~Time evolution of the wavepacket center of mass along the $x$ direction. The initial state is prepared to be $\ket{\psi \left( 0 \right)} \propto \sum_{y} \ket{x = L_{x}/2, y}$.}
	\label{fig: 2D}
\end{figure}

The action~(\ref{eq: TQFT-2D}) generally describes non-Hermitian topological phenomena in two dimensions.
The current density is
\begin{equation}
  j_{i} ( \bm{x}, E ) = \frac{1}{2\pi} \sum_{j} \varepsilon_{ij} \partial_{j}
  \theta ( \bm{x}, E ),
  	\label{eq: current-2D}
\end{equation}
which shows a particle flow in the direction perpendicular to the gradient of $\theta$.
Now, suppose that $\theta$ is spatially modulated along the $y$ direction.
Naively, such a $y$-dependent texture leads to a flow along the $y$ direction and is irrelevant to transport along the $x$ direction. 
However, Eq.~(\ref{eq: current-2D}) describes a perpendicular flow along the $x$ direction as a result of non-Hermitian topology.
To confirm this phenomenon, we investigate the lattice model $H = H_{0} + V$ with 
$H_{0} (\bm{k}) = \sigma_{x} \sin k_{x} + \sigma_{y} \sin k_{y} + \ii \gamma \left( \cos k_{x} + \cos k_{y} \right)$
and 
$V (\bm{x}) = \sigma_{z} \sin \phase (\bm{x}) + \ii \gamma \cos \phase (\bm{x})$.
Here, $\phase$ is given as $\phase (\bm{x}) = \pi/2 - 2\pi\texture y/L_{y}$, leading to nontrivial $\theta$~\cite{supplement}.
While no topological features appear in the absence of the spatial texture, the $y$-dependent texture induces the complex-spectral winding [Fig.~\ref{fig: 2D}\,(a)].
Consequently, a particle flow along the $x$ direction arises [Fig.~\ref{fig: 2D}\,(b)], which is consistent with the topological field theory description~(\ref{eq: current-2D}).
This perpendicular transport accompanies the perpendicular skin effect under the open boundary conditions~\cite{supplement}.
The number of the skin modes is controlled by the spatial gradient $\texture$, which is also a unique feature of two-dimensional systems.

\paragraph{Discussions.\,---}
In this Letter, we develop topological field theory of non-Hermitian systems.
Because of the dissipative and nonequilibrium nature of non-Hermiticity, the
temporal degree of freedom is distinguished from the spatial degrees of freedom,
and the field theory is formulated solely by the latter.
This theory provides the universal understanding of non-Hermitian topological phenomena.
We also demonstrate that the non-Hermitian skin effect originates from an anomaly.
For Hermitian systems, topological field theory is relevant not only to noninteracting systems but also to disordered and interacting systems. Similarly, our theory should be applicable to non-Hermitian systems with disorder and interaction.
Finally, it is noteworthy that other types of nonequilibrium topological field theory have recently been developed for Floquet operators~\cite{Glorioso-19} and Lindblad master equations~\cite{Tonielli-20}.

\medskip
K.K. thanks Hosho Katsura, Ryohei Kobayashi, Yasunori Lee, and Masahito Ueda for helpful discussions. K.K. and K.S. thank Takumi Bessho for helpful discussions. K.K. is supported by KAKENHI Grant No.~JP19J21927 from the Japan Society for the Promotion of Science (JSPS).
K.S. is supported by JST CREST Grant No.~JPMJCR19T2 and JST PRESTO Grant No.~JPMJPR18L4.
S.R. is supported by the National Science Foundation under award number DMR-2001181, and by a Simons Investigator Grant from the Simons Foundation (Award Number: 566116).

\bibliography{NH-topo}

\widetext
\pagebreak

\renewcommand{\theequation}{S\arabic{equation}}
\renewcommand{\thefigure}{S\arabic{figure}}
\renewcommand{\thetable}{S\arabic{table}}
\setcounter{equation}{0}
\setcounter{figure}{0}
\setcounter{table}{0}

\begin{center}
{\bf \large Supplemental Material for 
``Topological Field Theory of Non-Hermitian Systems"}
\end{center}

\section{Skin effect of non-Hermitian Dirac Hamiltonians}

Using the non-Hermitian Dirac Hamiltonian
\begin{equation}
H \left( k \right) = k + \ii \gamma,
\end{equation}
we discuss the non-Hermitian skin effect in one dimension. Here, $\gamma \in \mathbb{R}$ denotes the degree of non-Hermiticity. The corresponding topological invariant is given as the winding number
\begin{equation}
\WEone := - \oint_{\rm BZ} \frac{dk}{2\pi\ii} \left( \frac{d}{dk} \log \det \left[ H \left( k \right) - E \right] \right)
= - \int_{-\infty}^{\infty} \frac{dk}{2\pi\ii} \left( \frac{d}{dk} \log \left( k + \ii \gamma - E \right) \right)
= \frac{\mathrm{sgn} \left( \gamma - \mathrm{Im}\,E \right)}{2}.
\end{equation}

We consider an infinite system with a domain wall $x=0$, at which the topological invariant $\WEone$ changes. The Hamiltonian reads
\begin{equation}
H \left( x \right) = \begin{cases}
-\ii \partial_{x} + \ii \gamma & \mathrm{for}~~~x<0, \\
-\ii \partial_{x} - \ii \gamma & \mathrm{for}~~~x>0,
\end{cases}
\end{equation}
where $\gamma > 0$ is assumed. Let $E \in \mathbb{C}$ be an eigenenergy of $H \left( x \right)$, and $\varphi \left( x \right) \in \mathbb{C}$ be the corresponding eigenstate. For $x<0$, the Schr\"odinger equation reads
\begin{equation}
-\ii \frac{d\varphi}{dx} + \ii \gamma \varphi = E \varphi,
\end{equation}
leading to 
\begin{equation}
\varphi \left( x \right) = \varphi \left( 0 \right) e^{\left( \gamma + \ii E \right) x}.
\end{equation}
This eigenstate is localized at the domain wall $x=0$ with the localization length $\left( \gamma - \mathrm{Im}\,E \right)^{-1}$. For normalization of this eigenstate, we need
\begin{equation}
\gamma - \mathrm{Im}\,E > 0.
\end{equation}
Similarly, for $x > 0$, the eigenstate is given as
\begin{equation}
\varphi \left( x \right) = \varphi \left( 0 \right) e^{- \left( \gamma - \ii E \right) x},
\end{equation}
and the normalization condition reduces to
\begin{equation}
\gamma - \mathrm{Im}\,E > 0.
\end{equation}
Thus, the eigenenergy $E$ needs to satisfy
\begin{equation}
\left| \mathrm{Im}\,E \right| < \gamma 
\end{equation}
so that the localized eigenstate $\varphi \left( x \right)$ can be normalized. Importantly, an infinite number of $E$ can satisfy this condition. Consequently, an infinite number of boundary states appear, i.e., non-Hermitian skin effect. Such anomalous localization is not allowed in Hermitian systems, in which only symmetry-protected modes with $E=0$ can appear at the domain wall.

\section{Unidirectional transport in one dimension}

We describe unidirectional dynamics of non-Hermitian systems in one dimension. We here focus on the single-particle dynamics of a non-Hermitian system with a single band. Let
\begin{equation}
\ket{\psi \left( t \right)} = \sum_{x=1}^{L} c \left( x, t \right) \ket{x} 
\end{equation}
be the wavepacket at time $t$. Then, the center of mass of this wavepacket is 
\begin{equation}
\braket{x \left( t \right)} 
= \frac{\braket{\psi \left( t \right) |\,x\,| \psi \left( t \right)}}{\braket{\psi \left( t \right) | \psi \left( t \right)}}
= \frac{\sum_{x=1}^{L} x \left| c \left( x, t \right) \right|^{2}}{\sum_{x=1}^{L} \left| c \left( x, t \right) \right|^{2}}.
	\label{Seq: COM-1D}
\end{equation}
In momentum space, the Hamiltonian is diagonal, and the wavepacket is given as
\begin{equation}
\ket{\psi \left( t \right)} = \sum_{k} \tilde{c} \left( k \right) e^{-\ii E \left( k \right) t} \ket{k},\quad
k \in \left\{ 0, \frac{2\pi}{L}, \frac{4\pi}{L}, \cdots, \frac{2 \left( L-1 \right) \pi}{L} \right\},
\end{equation}
where
\begin{equation}
\ket{k} := \frac{1}{\sqrt{L}} \sum_{x=1}^{L} e^{\ii kx} \ket{x}
\end{equation}
is the momentum basis, and $\tilde{c} \left( k \right)$ is given by the initial state as
\begin{equation}
\tilde{c} \left( k \right) := \braket{k | \psi \left( 0 \right)}
= \frac{1}{\sqrt{L}} \sum_{x=1}^{L} c \left( x, 0 \right) e^{-\ii k x}.
\end{equation}
Then, we have
\begin{equation}
c \left( x, t \right) = \braket{x | \psi \left( t \right)}
= \frac{1}{\sqrt{L}} \sum_{k} \tilde{c} \left( k \right) e^{-\ii E \left( k \right) t + \ii k x},
\end{equation}
by which we can obtain $\braket{x \left( t \right)}$ in Eq.~(\ref{Seq: COM-1D}).

\begin{figure}[H]
\centering
\includegraphics[width=75mm]{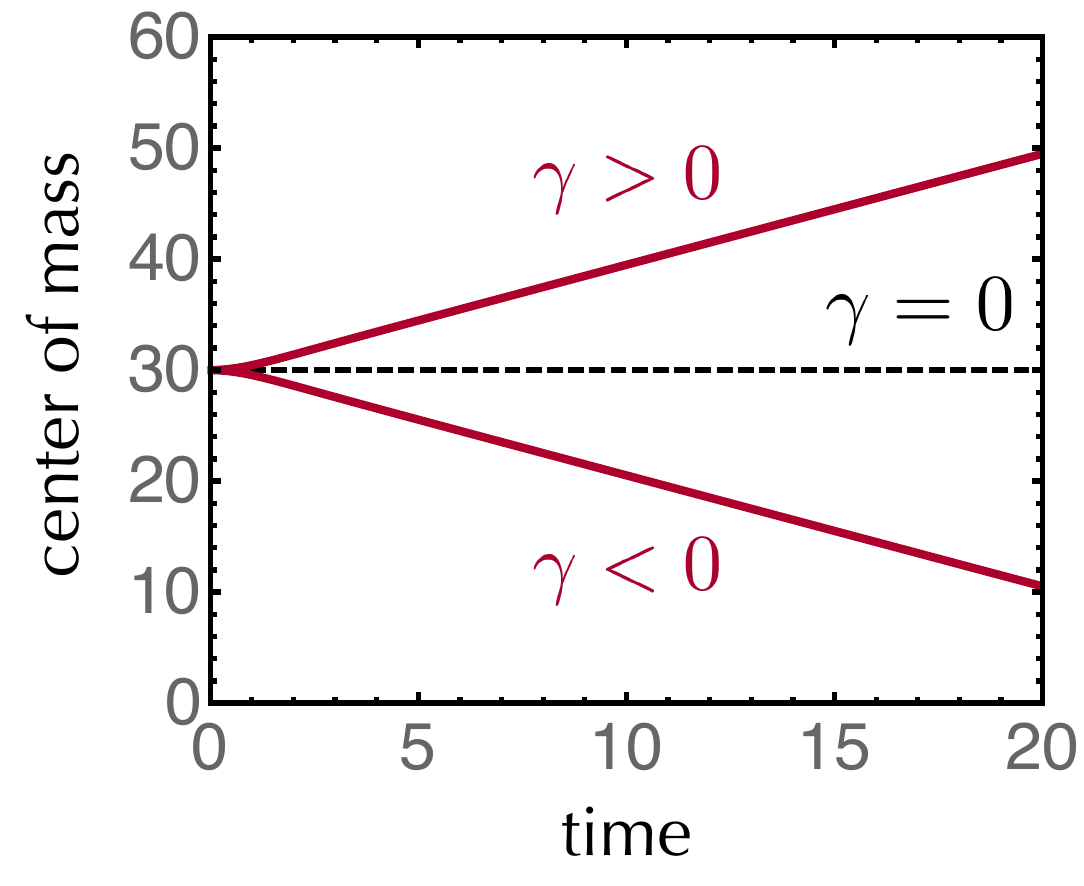} 
\caption{Unidirectional transport in the Hatano-Nelson model. The time evolution of the wavepacket center of mass is shown for the Hermitian case ($\gamma = 0$, black dashed line) and the non-Hermitian case ($\gamma = \pm 0.5$, red solid lines). The periodic boundary conditions are imposed with $L = 60$. The initial state is chosen as the single-site excitation $\ket{\psi \left( 0 \right)} = \ket{x_{0}}$ with $x_{0} = L/2 = 30$. While the wavepacket does not move in the Hermitian case, it moves unidirectionally in the non-Hermitian case. The direction of the unidirectional transport depends on the sign of the non-Hermiticity $\gamma$.} 
	\label{Sfig: HN-dynamics}
\end{figure}

Figure~\ref{Sfig: HN-dynamics} shows the time evolution of the wavepacket center of mass in the Hatano-Nelson model~\cite{Hatano-Nelson-96, *Hatano-Nelson-97}, whose spectrum reads
\begin{equation}
E ( k ) = -\cos k + \ii \gamma \sin k, 
\end{equation}
where $\gamma \in \mathbb{R}$ denotes the degree of non-Hermiticity. While the wavepacket does not move in the Hermitian case, it moves unidirectionally in the non-Hermitian case. 
The velocity of the wavepacket coincides with the group velocity $\partial_{k} \mathrm{Re} E (k) = \sin k$ for $k = \mathrm{sgn} (\gamma)\,(\pi/2)$, where the direction of the wavepacket depends on the sign of the non-Hermiticity $\gamma$. These results are consistent with the topological field theory description developed in the main text, which describes the unidirectional particle flow $j ( x ) = W_{1}$.

\section{Chiral anomaly}
	
In three dimensions, a non-Hermitian Dirac Hamiltonian is given, for example, as 
\begin{equation}
H \left( \bm{k} \right) = k_{x} \sigma_{x} - \ii k_{y} + k_{z} \sigma_{y} + \gamma \sigma_{z},
	\label{seq: NH-Dirac-3D}
\end{equation}
with $\gamma \in \mathbb{R}$. This Hamiltonian does not respect nontrivial symmetry. The topological invariant is given as the winding number~\cite{KSUS-19} 
\begin{equation}
W_{3} := - \sum_{ijk} \oint_{\rm BZ} \frac{d^{3}k}{24\pi^{2}}\,\varepsilon_{ijk}\,\mathrm{tr} \left[ \left( H^{-1} \left( \bm{k} \right) \frac{\partial H \left( \bm{k} \right)}{\partial k_{i}}\right) \left( H^{-1} \left( \bm{k} \right) \frac{\partial H \left( \bm{k} \right)}{\partial k_{j}}\right) \left( H^{-1} \left( \bm{k} \right) \frac{\partial H \left( \bm{k} \right)}{\partial k_{k}}\right)\right]
= \frac{\mathrm{sgn} \left( \gamma \right)}{2},
\end{equation}
where reference energy is assumed to be zero for simplicity. Notably, $H \left( \bm{k} \right)$ consists of the Hermitian Dirac Hamiltonian $k_{x} \sigma_{x} + k_{z} \sigma_{y} + \gamma \sigma_{z}$ for a Chern insulator and the scalar imaginary term $-\ii k_{y}$. The topological invariant $W_{3}$ coincides with the Chern number of $k_{x} \sigma_{x} + k_{z} \sigma_{y} + \gamma \sigma_{z}$.

In the presence of a boundary, the topological field theory has an anomaly, and additional degrees of freedom are required at the boundary to retain gauge invariance. Suppose that the open boundary conditions are imposed along the $z$ direction. Here, the Chern insulator $k_{x} \sigma_{x} + k_{z} \sigma_{y} + \gamma \sigma_{z}$ hosts a chiral fermion $\pm k_{x}$ at the boundary. Thus, the boundary modes of the non-Hermitian Dirac Hamiltonian in Eq.~(\ref{seq: NH-Dirac-3D}) are given as
\begin{equation}
\HB \left( \bm{k} \right) = \pm k_{x} - \ii k_{y}.
\label{eq:bdy_mode}
\end{equation}
The point gap with $E=0$ closes at $\bm{k} = 0$, around which the winding number $W_{1}$ is nontrivial. 

The corresponding action reads
\begin{eqnarray}
S^{\rm boundary} \left[ \bm{A} \left( \bm{x} \right) \right]
&=& \int \left[ \bar{\psi}_{\rm R} \left( -\ii \partial_{x} - A_{x} \right) \psi_{\rm R} + \bar{\psi}_{\rm R} \left( -\partial_{y} - \ii A_{y} \right) \psi_{\rm R} \right. \nonumber \\
&&\qquad\qquad\qquad \left. + \bar{\psi}_{\rm L} \left( +\ii \partial_{x} + A_{x} \right) \psi_{\rm L} + \bar{\psi}_{\rm L} \left( -\partial_{y} - \ii A_{y} \right) \psi_{\rm L} \right] dxdy.
\end{eqnarray}
Here, $\psi_{\rm R}$ and $\bar{\psi}_{\rm R}$ denote the matter degrees of freedom at the surface characterized by $H \left( \bm{k} \right) = k_{x} - \ii k_{y}$, while $\psi_{\rm L}$ and $\bar{\psi}_{\rm L}$ at the other surface characterized by $H \left( \bm{k} \right) = -k_{x} - \ii k_{y}$. Importantly, $S^{\rm boundary}$ reduces to the action of chiral fermions in $1+1$ dimensions by replacing $y$ with imaginary time $\ii t$. Consequently, $S^{\rm boundary}$ has a chiral anomaly~\cite{Peskin-textbook}, which compensates for the anomaly of the bulk. The anomaly equation reads
\begin{equation}
\sum_{i \in \left\{ x, y \right\}} \partial_{i} j_{i}^{A}
= \frac{1}{\pi} \sum_{ij \in \left\{ x, y \right\}} \varepsilon_{ij} \partial_{i} A_{j}
= \frac{B}{\pi},
	\label{seq: chiral anomaly}
\end{equation}
where $B := \partial_{x} A_{y} - \partial_{y} A_{x}$ is a magnetic field along the $z$ direction, and the axial current $\bm{j}^{A}$ is defined as
\begin{equation}
j_{x}^{A} := \bar{\psi}_{\rm R} \psi_{\rm R} + \bar{\psi}_{\rm L} \psi_{\rm L},\quad
j_{y}^{A} := \bar{\psi}_{\rm R} \psi_{\rm R} - \bar{\psi}_{\rm L} \psi_{\rm L}.
\end{equation}
The corresponding (Noether) charge is $N_{\rm R} - N_{\rm L}$ with
\begin{equation}
N_{\rm R} := \int \bar{\psi}_{\rm R} \psi_{\rm R}\,dx,\quad
N_{\rm L} := \int \bar{\psi}_{\rm L} \psi_{\rm L}\,dx.
\end{equation}
With the global quantities $N_{\rm R}$ and $N_{\rm L}$, as well as the magnetic flux $\Phi := \int B\,dxdy$ on the surface, the anomaly equation~(\ref{seq: chiral anomaly}) reduces to
\begin{equation}
N_{\rm R} - N_{\rm L} = \sum_{i} \int \left( \partial_{i} j_{i}^{A} \right) dxdy
= \frac{\Phi}{\pi}.
\end{equation}
On the other hand, $\mathrm{U} \left( 1 \right)$ symmetry leads to $N_{\rm R} + N_{\rm L} = 0$. Hence, we have
\begin{equation}
N_{\rm R} = \frac{\Phi}{2\pi},\quad
N_{\rm L} = - \frac{\Phi}{2\pi}.
\end{equation}
Thus, the charges $N_{\rm R}$ and $N_{\rm L}$ appear at the surfaces perpendicular to the magnetic field. Similarly to the one-dimensional case, these charges correspond to skin modes. The number of these skin modes is proportional to the number of the fluxes.

\begin{figure}[H]
\centering
\includegraphics[width=100mm]{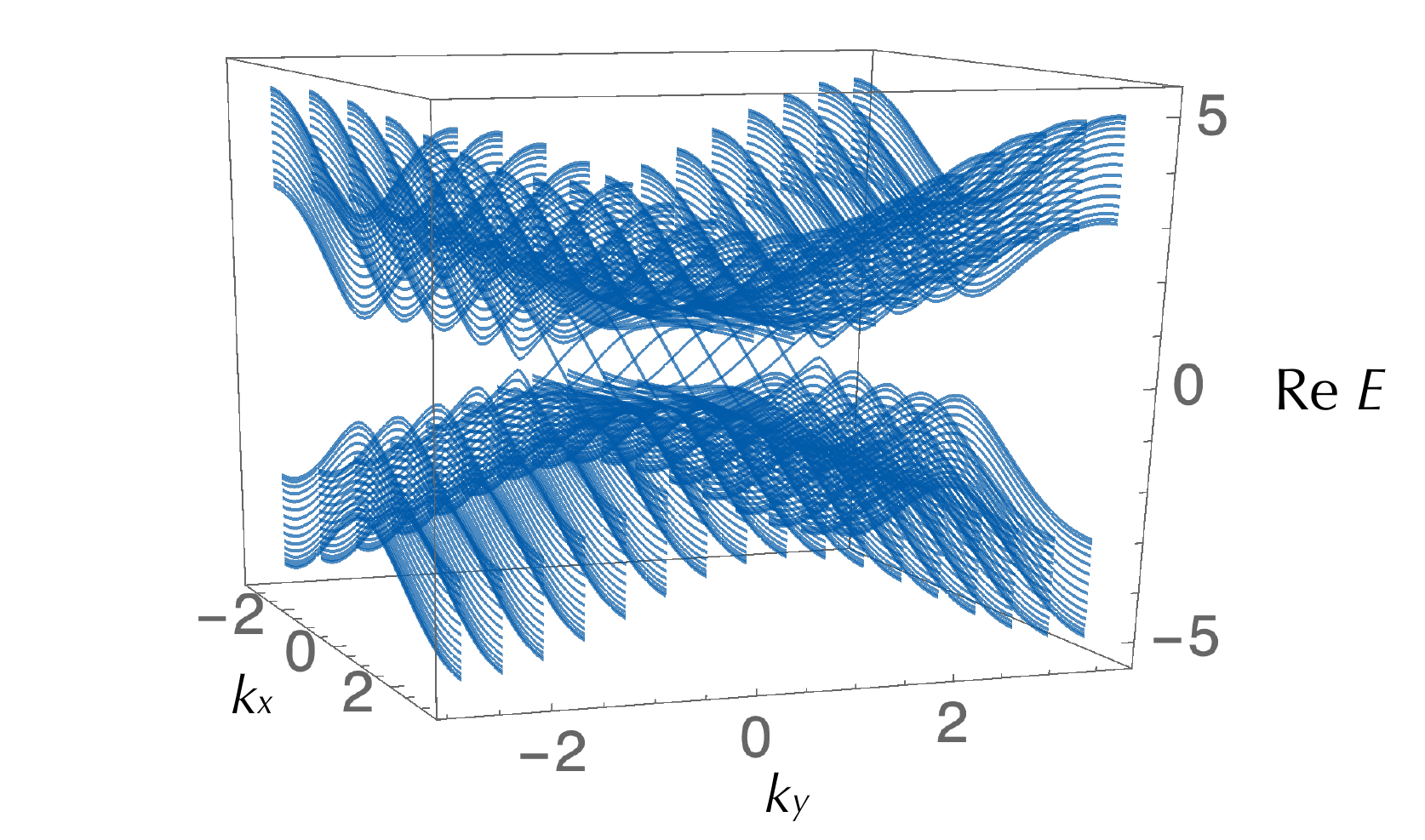} 
\caption{Real part of the spectrum of the non-Hermitian Hamiltonian~\eqref{seq: lattice_model} with $\gamma = 1$ and $L_{z} = 15$. The periodic boundary conditions are imposed along the $x$ and $y$ directions, while the open boundary conditions are imposed along the $z$ direction. The spectra of some slices of $k_{y}$'s are shown.}
	\label{Sfig: surf}
\end{figure}

In addition, we provide a lattice model that exhibits the surface states with the dispersion relation~\eqref{eq:bdy_mode}. 
A three-dimensional lattice model that reduces to the low-energy theory~\eqref{seq: NH-Dirac-3D} around $\bm{k} = 0$ is given as
\begin{align}
    H(\bm{k})
    =
    \sigma_x \sin k_x - \ii \sin k_y + \sigma_y \sin k_z + \sigma_z \left( \gamma - 3 + \cos k_x + \cos k_y + \cos k_z \right)
	\label{seq: lattice_model}
\end{align}
in momentum space. 
The Hermitian part of this model describes a Weyl semimetal. 
Under the open boundary conditions along the $z$ direction, this hosts the surface modes if and only if $\gamma - 2 + \cos k_x + \cos k_y$ and $\gamma - 4 + \cos k_x + \cos k_y$ have opposite signs. 
Let us focus on the parameter region $0 < \gamma < 2$ so that a single surface mode will appear in the momenta such that $\cos k_x+\cos k_y>2-\gamma$ is satisfied within a disc including the $\Gamma$ point $(k_x,k_y) = (0,0)$.
On the top (bottom) boundary, the orbital degrees of freedom of the surface mode have the chirality $\sigma_x=1$ ($\sigma_x=-1$), which means the dispersion relation
\begin{equation}
H^{\rm top} (k_x, k_y)  = \sin k_x - \ii \sin k_y\quad
\left[ H^{\rm bottom} (k_x, k_y) = -\sin k_x - \ii \sin k_y \right].
\end{equation}
This indeed reduces to Eq.~\eqref{eq:bdy_mode} near $(k_x,k_y)=(0,0)$. 
Figure~\ref{Sfig: surf} shows the numerically-obtained spectrum of the surface modes in the topological phase.

\section{Chiral magnetic skin effect}

As discussed in the main text,
a chiral anomaly of $\left( 2+0 \right)$-dimensional boundary states results
in the chiral magnetic skin effect.
This result can also be confirmed in a lattice model.
Here, we investigate the following non-Hermitian system in three dimensions:
\begin{equation}
H \left( \bm{k} \right) 
= \cos k_{x} + \cos k_{y} + \cos k_{z} + \ii \gamma \left( \sigma_{x} \sin k_{x} + \sigma_{y} \sin k_{y} + \sigma_{z} \sin k_{z} \right),
	\label{Seq: 3D-Hamiltonian}
\end{equation}
where $\gamma \in \mathbb{R}$ denotes the degree of non-Hermiticity. This system is characterized by the three-dimensional winding number $W_{3}$ in terms of a point gap.
Notably, $W_{3}$ is invariant under the change of the Hamiltonian $H \left( \bm{k} \right) \mapsto U H \left( \bm{k} \right) V^{-1}$ with independent unitary matrices $U$ and $V$. The Hamiltonian in Eq.~(\ref{Seq: 3D-Hamiltonian}) reduces to the Hamiltonian in Eq.~(\ref{seq: lattice_model}) by the transformation $H \left( \bm{k} \right) \mapsto H \left( \bm{k} \right) \sigma_{z}$, and hence $W_{3}$ is invariant. Thus, consequent topological phenomena, including the chiral magnetic skin effect, are expected to occur regardless of the specific choice of the Hamiltonian.

Let us impose a magnetic field along the $z$ direction. In particular, we consider the vector potential
\begin{equation}
A_{x} = - \frac{2\pi my}{L_{x} L_{y}},\quad
A_{y} = \begin{cases}
~~~~0 & \left( y = 1, \cdots, L-1 \right), \\
\cfrac{2\pi mx}{L_{x}} & \left( y = L \right),
\end{cases}\quad
A_{z} = 0,
	\label{Seq: 3D-vector-potential}
\end{equation}
where $m$ is an integer, and $L_{x}$ and $L_{y}$ are the lengths of the system along the $x$ and $y$ directions, respectively. This vector potential gives
\begin{equation}
\int B_{z} dx dy = 2\pi m,
\end{equation}
which means the presence of $m$ magnetic fluxes (note that $2\pi$ is the flux quantum in the natural units).

\begin{figure}[H]
\centering
\includegraphics[width=172mm]{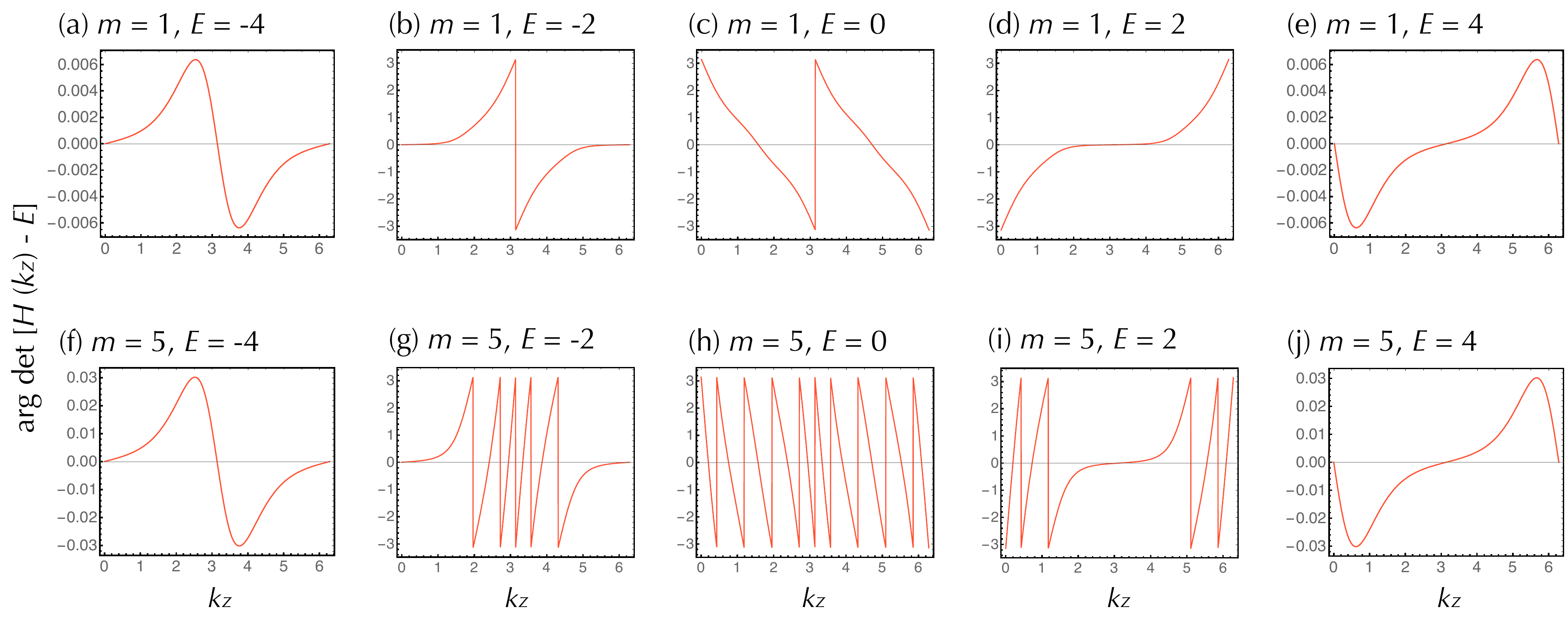} 
\caption{Winding number $\WEone$ in the presence of magnetic fluxes. For the non-Hermitian Hamiltonian~(\ref{Seq: 3D-Hamiltonian}) with the vector potential~(\ref{Seq: 3D-vector-potential}), the argument of $\det \left[ H \left( k_{z} \right) - E \right]$ is shown as a function of the wavenumber $k_{z}$. The parameters are chosen as $\gamma = 0.5$ and $L_{x} = L_{y} = 10$. The periodic boundary conditions are imposed along all the directions. The number of the magnetic fluxes is $m = 1$ for (a-e), and $m=5$ for (f-j). The reference energy is $E = -4$ for (a, f), $E = -2$ for (b, g), $E=0$ for (c, h), $E = 2$ for (d, i), and $E = 4$ for (e, j). The winding numbers are given as $\WEone = 0$ for $E = \pm 4$ (a, e, f, j), $\WEone = -m$ for (b, d, g, i), and $\WEone = 2m$ for (c, h).}
	\label{Sfig: 3D-winding}
\end{figure}

In the presence of a magnetic field, the three-dimensional winding number of $H \left( \bm{k} \right)$ leads to the one-dimensional winding number along the direction 
parallel
to the magnetic field. In particular, if a magnetic field is imposed along the $z$ direction, the one-dimensional winding number
\begin{equation}
\WEone = - \oint_{0}^{2\pi} \frac{dk_z}{2\pi} \left( \frac{d}{dk_z} \mathrm{arg} \det \left[ H \left( k_{z} \right) - E \right] \right)
\end{equation}
can be nontrivial. Here, $\mathrm{arg}$ denotes the argument of a complex number. Figure~\ref{Sfig: 3D-winding} shows the winding numbers $\WEone$ for various choices of the reference energy $E$ and the number $m$ of magnetic fluxes. While $\WEone$ vanishes for $E$ such that the three-dimensional winding number $\WEthree$ vanishes, $\WEone$ is nonzero for $E$ such that $\WEthree$ is nonzero. Notably, $\WEone$ is proportional to $m$.

\begin{figure}[H]
\centering
\includegraphics[width=172mm]{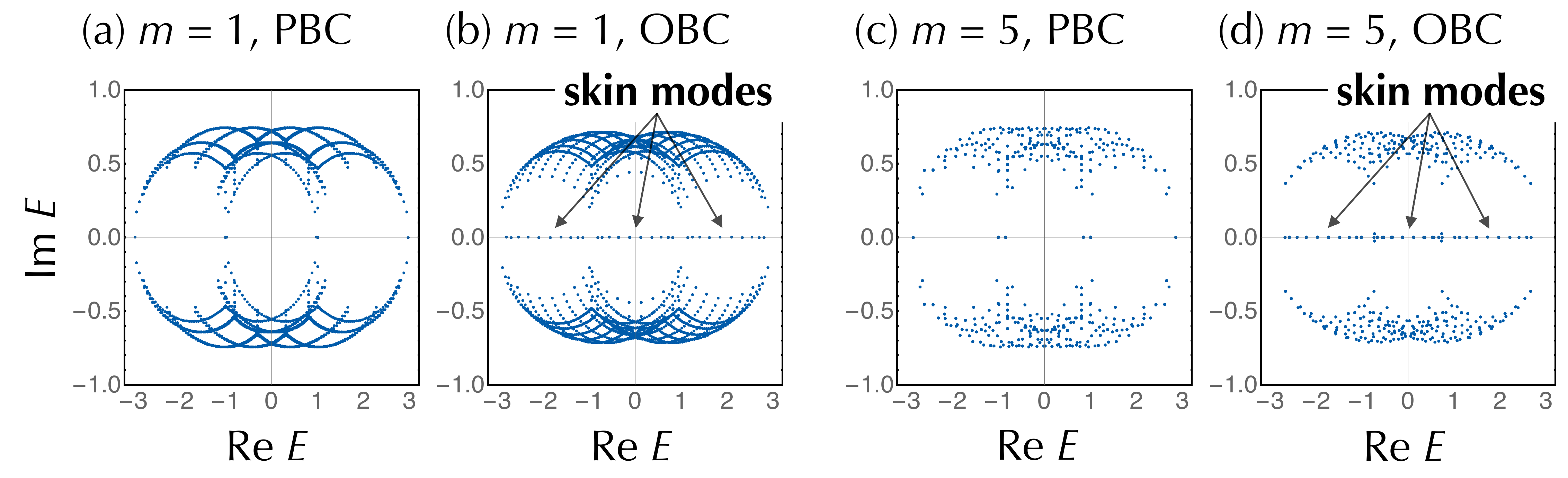} 
\caption{Chiral magnetic skin effect. Complex spectra of the non-Hermitian Hamiltonian~(\ref{Seq: 3D-Hamiltonian}) with the vector potential~(\ref{Seq: 3D-vector-potential}) are shown. The parameters are chosen as $\gamma = 0.5$ and $L_{x} = L_{y} = L_{z} = 10$. The number of magnetic fluxes is $m=1$ for (a, b), and $m=5$ for (c, d). The periodic boundary conditions are imposed along the $x$ and $y$ directions. Along the $z$ direction, the periodic boundary conditions are imposed for (a, c), and the open boundary conditions are imposed for (b, d). Skin modes appear under the open boundary conditions (b, d).}
	\label{Sfig: 3D-skin}
\end{figure}

According to the field-theoretical discussions developed in the main text, skin modes generally appear at a boundary perpendicular to a magnetic field when the three-dimensional system supports the nontrivial winding number. Consistently, under the magnetic field along the $z$ direction, skin modes appear under the open boundary conditions along the $z$ direction (Fig.~\ref{Sfig: 3D-skin}).
A similar result has recently been reported in Ref.~\cite{Nakamura-20}.
This is also consistent with the nontrivial (one-dimensional) winding number shown in Fig.~\ref{Sfig: 3D-winding}. We note that the number of skin modes is not necessarily proportional to the number of magnetic fluxes in finite systems. The field-theoretical discussions assume the infinite degrees of freedom and do not strictly apply to finite systems. Nevertheless, they explain qualitative behavior of non-Hermitian topological phenomena even in finite systems, as shown here for the chiral magnetic skin effect. Similarly, in one dimension, the number of skin modes is not necessarily proportional to the winding number $W_{1}$ in finite systems, although we strictly have this proportional relationship in semi-infinite systems~\cite{OKSS-20}.

\section{Transport and skin effect induced by a perpendicular spatial texture}

As described in the main text, the effective action of non-Hermitian systems in two dimensions is generally  
\begin{equation}
S [ \bm{A} ]
  = \frac{1}{2\pi} \sum_{ij} \int \theta (\bm{x}) \varepsilon_{ij}
      \partial_{i} A_{j} ( \bm{x}) d^{2}x
  = \frac{1}{2\pi} \int \theta (\bm{x}) B ( \bm{x}) d^{2}x,
	\label{Seq: TQFT-2D}
\end{equation}
where $B := \sum_{ij} \varepsilon_{ij} \partial_{i} A_{j}$ is the magnetic field. Here, $\theta$ can be identified as the Wess-Zumino term~\cite{WZ-71}:
\begin{equation}
\theta = - \frac{1}{24\pi^2} \oint_{\mathrm{BZ} \times [0,1]} \mathrm{tr} [ H^{-1}d H]^3 \in \mathbb{R}/\mathbb{Z},
\end{equation}
where $\tilde{H} = \tilde H(\bm{k},s)$ is an extension of the two-dimensional Hamiltonian $H (\bm{k})$ that satisfies
\begin{equation}
\tilde H (\bm{k} ,s=0) = H(\bm{k}),\quad
\tilde H (\bm{k} ,s=1) = H_{0}
\end{equation}
with a constant Hamiltonian $H_{0}$. In other words, $\theta$ gives the integral of the three-dimensional winding number density for an extension $\tilde H(k_x,k_y,s)$. Here, for simplicity, we set reference energy $E$ to zero and assume the absence of the one-dimensional winding number. The Wess-Zumino term $\theta$ is a non-Hermitian analog of the electric polarization in $\left( 1+1 \right)$-dimensional Hermitian systems.

While Eq.~(\ref{Seq: TQFT-2D}) generally describes non-Hermitian systems in two dimensions, certain symmetry is needed for the quantization of $\theta$. For example, suppose that the system is invariant under reciprocity [i.e., $\mathcal{T} H^{T}( \bm{k} ) \mathcal{T}^{-1} = H ( \bm{k} )$ with a unitary matrix $\mathcal{T}$ satisfying $\mathcal{T}\mathcal{T}^{*} = - 1$; class AII$^{\dag}$~\cite{KSUS-19}]. In terms of reciprocity, a magnetic field $B$ is odd. Hence, $\theta$ should also be odd under reciprocity so that the action will be reciprocal. Since $\theta$ is well defined only modulo $1$, it is quantized to be the $\mathbb{Z}_{2}$ values $\theta = 0, 1/2$, which gives the $\mathbb{Z}_{2}$ classification of the topological phase. Moreover, $\theta$ is quantized also in spinless superconductors that respect particle-hole symmetry (class D~\cite{KSUS-19}). The $\mathbb{Z}_{2}$ quantization is reminiscent of the quantized polarization in particle-hole-symmetric topological insulators in $1+1$ dimensions and time-reversal-invariant topological insulators in $3+1$ dimensions~\cite{QHZ-08}.

\begin{figure}[b]
\centering
\includegraphics[width=144mm]{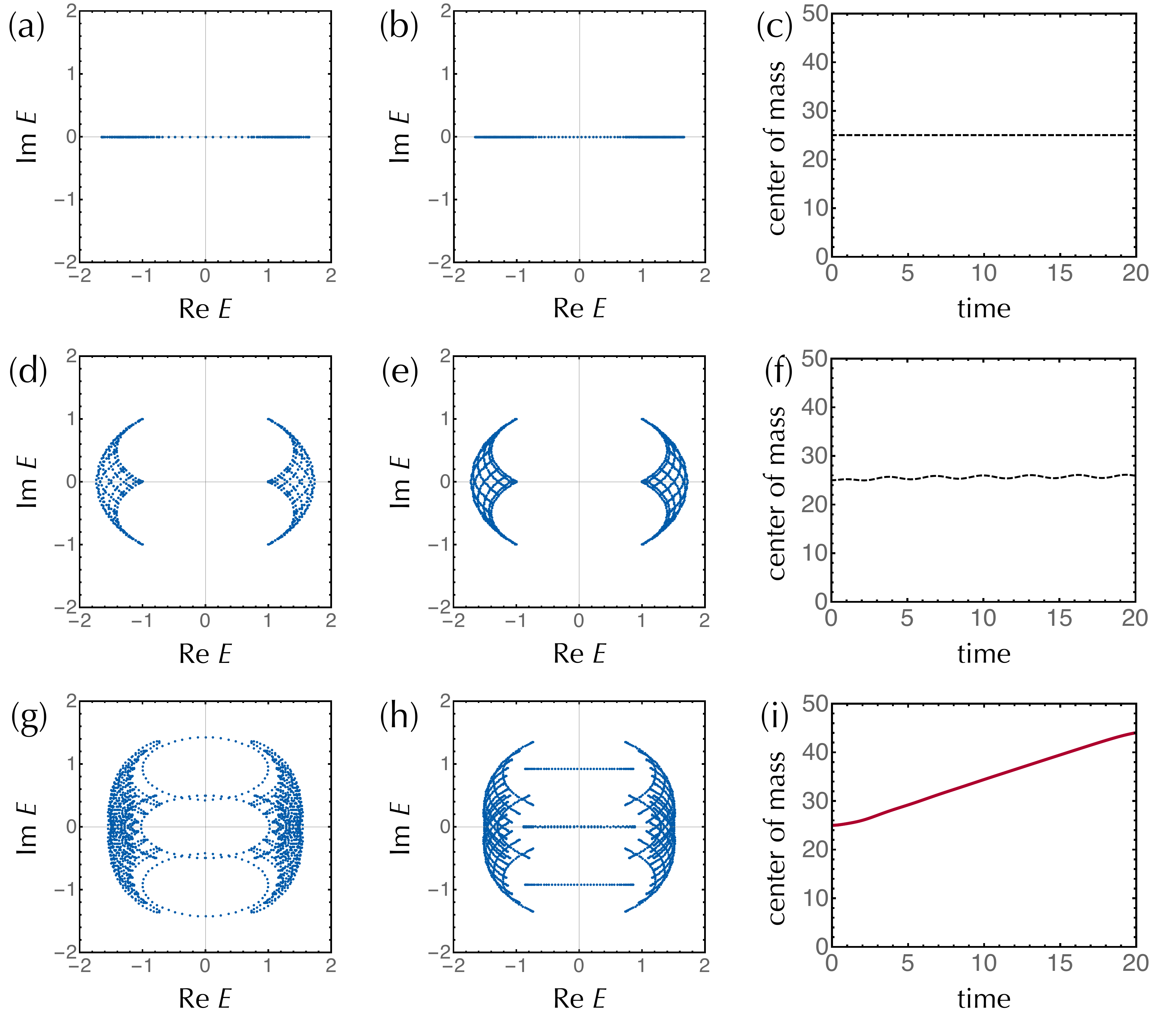} 
\caption{Non-Hermitian topological phenomena in two dimensions. Complex spectra of the non-Hermitian Hamiltonian~(\ref{Seq: NH-2D-Hamiltonian}) with $L_{x} = 50$ and $L_{y} = 20$ are shown under the periodic boundary conditions for~(a, d, g) and under the open boundary conditions along the $x$ direction for~(b, e, h). The corresponding time evolution of the wavepacket center of mass along the $x$ direction is shown for~(c, f, i), where the periodic boundary conditions are imposed, and the initial state is prepared to be $\ket{\psi \left( 0 \right)} \propto \sum_{y} \ket{x = L_{x}/2, y}$.  (a, b, c)~Hermitian system with the spatial texture ($\gamma = 0$ and $\texture=1$). (d, e, f)~Non-Hermitian system without the spatial texture ($\gamma = 0.5$ and $\texture=0$). (g, h, i)~Non-Hermitian system with the spatial texture ($\gamma = 0.5$ and $\texture=1$). Under the periodic boundary conditions, the complex spectrum is characterized by nontrivial winding numbers, which lead to the skin effect under the open boundary conditions. Consistently, the wavepacket center of mass moves unidirectionally.}
	\label{Sfig: 2D-spectrum-dynamics}
\end{figure}

\begin{figure}[t]
\centering
\includegraphics[width=172mm]{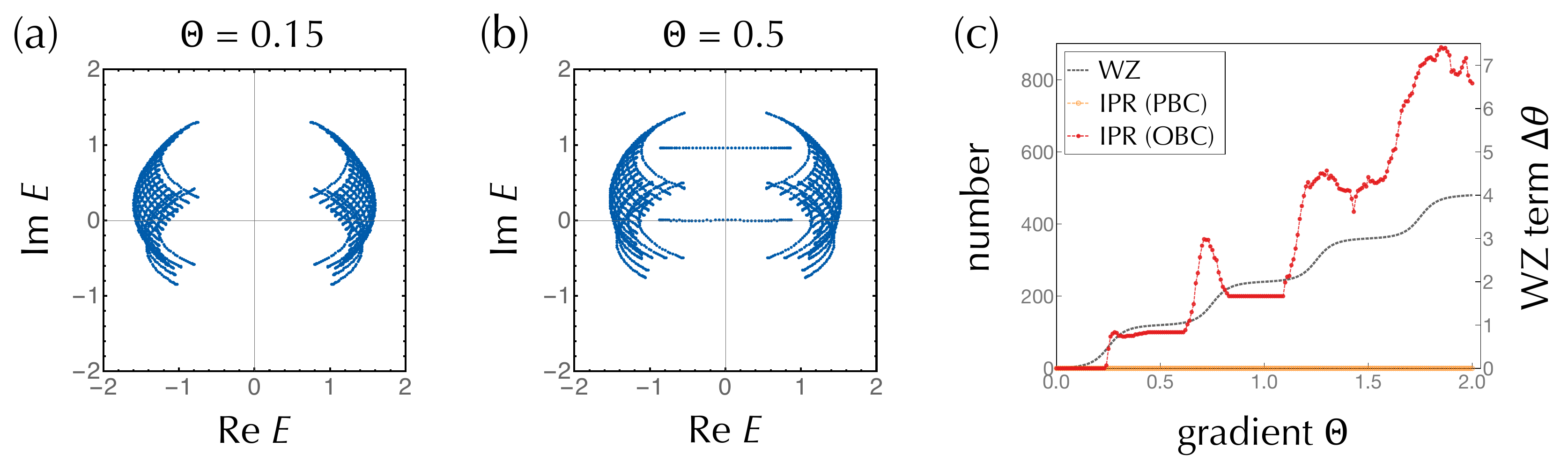} 
\caption{Non-Hermitian skin effect in two dimensions. Complex spectra and wavefunctions of the non-Hermitian Hamiltonian~(\ref{Seq: NH-2D-Hamiltonian}) with $L_{x} = 50$, $L_{y} = 20$, and $\gamma = 0.5$ are investigated for various choices of the spatial gradient $\texture$. The open boundary conditions are imposed along both $x$ and $y$ directions. (a)~Complex spectrum for $\texture = 0.15$. (b)~Complex spectrum for $\texture = 0.5$. (c)~Number of the skin modes. The localized eigenstates whose inverse participation ratios (IPRs) are less than $0.015$ are counted as a function of $\texture$ for the periodic (orange dots) and open (red dots) boundary conditions, which are compatible with the change of the Wess-Zumino (WZ) term along the $y$ direction [$\Delta \theta := \theta \left( y = L_{y} \right) - \theta \left( y = 0 \right)$; black curve]. The reference energy of the WZ term is set to zero.}
	\label{Sfig: 2D-skin}
\end{figure}

The effective action~(\ref{Seq: TQFT-2D}) describes phenomena induced by the spatial inhomogeneity of $\theta$. The corresponding current density is
\begin{equation}
j_{i} (\bm{x})
= \frac{\delta S [\bm{A}]}{\delta A_{i} (\bm{x})}
= \frac{1}{2\pi} \sum_{j} \varepsilon_{ij} \partial_{j} \theta (\bm{x}),
	\label{Seq: NH-2D-current}
\end{equation}
which shows a particle flow in a direction perpendicular to the spatial gradient of $\theta$.
To understand this non-Hermitian topological phenomenon, we here consider the two-dimensional non-Hermitian system $H = H_{0} + V$ with
\begin{align}
H_{0} (\bm{k}) 
= \sigma_{x} \sin k_{x} + \sigma_{y} \sin k_{y} + \ii \gamma \left( \cos k_{x} + \cos k_{y} \right),\qquad
V (\bm{x})
= \sigma_{z} \sin \phase (\bm{x}) + \ii \gamma \cos \phase (\bm{x}).
	\label{Seq: NH-2D-Hamiltonian}
\end{align}
We assume that $V$ is translationally invariant along the $x$ direction and modulated only along the $y$ direction. In particular, we consider
\begin{equation}
\phase (\bm{x}) = \frac{\pi}{2} - \frac{2\pi\texture}{L_{y}} y,
\end{equation}
where $\texture$ describes the spatial gradient of the texture along the $y$ direction. 
This spatial texture induces the nontrivial Wess-Zumino term $\theta$ [see Fig.~\ref{Sfig: 2D-skin}\,(c)], which in turn leads to
the unidirectional transport and the 
skin effect along the $x$ direction. For $\gamma = 0$, the spectrum is entirely real and no skin effect occurs even in the presence of the spatial texture [Fig.~\ref{Sfig: 2D-spectrum-dynamics}\,(a, b)]. 
The transport phenomenon is understood by the wavepacket dynamics in a similar manner to the one-dimensional case (see the section ``Unidirectional transport in one dimension" for details). Consistently with the absence of the complex-spectral winding and the skin effect, the center of mass of a wavepacket does not move under the time evolution [Fig.~\ref{Sfig: 2D-spectrum-dynamics}\,(c)]. For $\gamma \neq 0$, the spectrum generally becomes complex, but neither skin effect nor unidirectional transport arises in the absence of the spatial texture [Fig.~\ref{Sfig: 2D-spectrum-dynamics}\,(d, e, f)]. 
In the simultaneous presence of $\gamma$ and the spatial texture, on the other hand, the system exhibits the non-Hermitian topological phenomenon due to nontrivial $\theta$. Under the periodic boundary conditions, the spectrum is characterized by nontrivial winding numbers in the complex energy plane [Fig.~\ref{Sfig: 2D-spectrum-dynamics}\,(g)], which induce the skin effect under the open boundary conditions [Fig.~\ref{Sfig: 2D-spectrum-dynamics}\,(h)]. Consistently, the wavepacket center of mass moves unidirectionally [Fig.~\ref{Sfig: 2D-spectrum-dynamics}\,(i)]. Notably, this transport along the $x$ direction arises from the spatial texture along the $y$ direction, which is a unique feature of the two-dimensional case that is distinct from the one-dimensional case. The topological field theory~(\ref{Seq: TQFT-2D}) correctly describes this non-Hermitian topological phenomenon.

Moreover, Fig.~\ref{Sfig: 2D-skin} shows the skin effect for various choices of the spatial gradient $\texture$. For sufficiently small $\texture$ ($\lesssim 1/4$), no skin effect occurs [Fig.~\ref{Sfig: 2D-skin}\,(a)]. With increasing $\texture$, skin modes gradually appear [Fig.~\ref{Sfig: 2D-skin}\,(b)]. We count the skin modes as a function of $\texture$ [Fig.~\ref{Sfig: 2D-skin}\,(c)]. Here, we investigate the inverse participation ratios $\sum_{x, y} \left| \psi \left( x, y \right) \right|^{4} /\sum_{x, y} \left| \psi \left( x, y \right) \right|^{2}$ for all the eigenstates $\psi$. They decrease with $\propto 1/L_x L_y$ for delocalized eigenstates but remain to be approximately one for localized eigenstates, and hence measure the degree of localization. 
It is notable that the skin effect along the $y$ direction or Anderson localization may arise since the texture breaks translation invariance along the $y$ direction. Still, as long as the periodic boundary conditions are imposed along the $x$ direction, the inverse participation ratios are small for all the eigenstates even under the open boundary conditions along the $y$ direction. Thus, no localization phenomena occur except for the skin effect along the $x$ direction. 
The number of the skin modes is compatible with the change of the Wess-Zumino term $\theta$.
As demonstrated by these results, the spatial texture enables control of the skin modes, which is also a unique feature of two-dimensional systems. 

\end{document}